\documentclass[aps,prb,twocolumn,groupedaddress,citeautoscript]{revtex4}

\usepackage{graphicx}
\usepackage{amsmath}
\usepackage{times}

\graphicspath{{.}{./EPS/}}

\begin{document}

\title{Conductance oscillations in strongly correlated fractional
quantum Hall line junctions}

\author{U. Z\"ulicke}
\affiliation{Institut f\"ur Theoretische Festk\"orperphysik,
Universit\"at Karlsruhe, D-76128 Karlsruhe, Germany}
\affiliation{Institute of Fundamental Sciences, Massey University,
Private Bag 11~222, Palmerston North, New Zealand}\thanks{Present and
permanent address.}

\author{E. Shimshoni}
\affiliation{Department of Mathematics--Physics, 
University of Haifa at Oranim, Tivon 36006, Israel}

\date{\today}

\begin{abstract}

We present a detailed theory of transport through line junctions formed
by counterpropagating single--branch fractional--quantum--Hall edge
channels having different filling factors. Intriguing transport
properties are exhibited when strong Coulomb interactions between
electrons from the two edges are present. Such strongly correlated line
junctions can be classified according to the value of an effective
line--junction filling factor $\tilde\nu$ that is the inverse of an even
integer. Interactions turn out to affect transport most importantly for
$\tilde\nu=1/2$ and $\tilde\nu=1/4$. A particularly interesting case is
$\tilde\nu=1/4$ corresponding to, e.g., a junction of edge channels
having filling factor $1$ and $1/5$, respectively. We predict its
differential tunneling conductance to oscillate as a function of
voltage. This behavior directly reflects the existence of novel
Majorana--fermion quasiparticle excitations in this type of line
junction. Experimental accessibility of such systems in current
cleaved--edge overgrown samples enables direct testing of our
theoretical predictions.

\end{abstract}


\maketitle

\section{Introduction}

One--dimensional (1D) electron systems~\cite{voit:reprog:94} have long
been the focus of theoretical and experimental research. Initially,
theorists studied them as rare examples of exactly soluble interacting
many--body systems~\cite{tom:prog:50,lutt:jmp:63,matt:jmp:65}. They
served as a basis for the development of powerful new theoretical tools
such as bosonization~\cite{schott:pr:69,lut:prb:74,kopietz} and
refermionization~\cite{vondelft} techniques. Due to their intriguing
non--Fermi--liquid properties, interacting 1D electron systems are
classified within the distinct phenomenology of {\em
Luttinger--liquid\/} behavior~\cite{fdmh:jpc:81}. Eventually,
realizations of quasi--1D electron systems were found in metallic
materials with strongly anisotropic resistivity~\cite{early1D}. Recent
fabrication of clean long semiconductor quantum wires~\cite{yac:prl:96}
as well as carbon nano\-tubes~\cite{dresselhaus} created new
possibilities to observe Luttinger--liquid behavior in experiment. 

An especially versatile type of 1D electron system is realized at the
boundary of two--dimensional (2D) electron systems in a strong
perpendicular magnetic field. At particular values of the filling factor
$\nu=2\pi\ell^2 n_{\text{2D}}$, where $\ell=\sqrt{\hbar c/|e B|}$ is the
magnetic length and $n_{\text{2D}}$ the electron sheet density, the 2D
system becomes incompressible in the bulk~\cite{ahmintro}, giving rise
to quantized values of the Hall resistance. In this regime where the
quantum Hall (QH) effect~\cite{qhe-sg,qhepersp} is observed, low--lying
excitations exist at the sample boundary~\cite{bih:prb:82} whose
electronic properties are analogous to chiral versions of Luttinger
liquids when the filling factor at the incompressibility is
fractional~\cite{wen:int:92}. Unlike the more conventional types of
quasi--1D metals, the properties of edge excitations in a QH sample can
be easily tailored. Simple adjustment of the magnetic field can create
different QH states in the bulk of the 2D system with concomitant change
in the edge's chiral--Luttinger--liquid properties. Advanced
nanostructuring techniques enable the creation of novel tunneling
geometries~\cite{amc:prl:96,michi:physe:01,matttunspec,mattcorner1}
involving QH edges as well as line junctions between
them~\cite{kang:nat:99,lorke:prb:02,mattcorner2}. While line junctions
between counterpropagating chiral edge channels having the same integer
filling factor closely mirror properties of conventional quasi--1D
systems~\cite{MG,KS,KF}, an entirely new arena for novel correlation
effects is opened up when the edge channels forming the junction belong
to fractional--QH samples. For the case of a disordered junction, these
have been investigated in Refs.~[\onlinecite{qhlinej1,qhlinej2}]. A
related study for a clean system of tunnel--coupled {\em
copropagating\/} fractional--QH edge channels in bilayers was performed
recently~\cite{naud:nucl:00,naud:prb:01}.

Here we consider the situation where the single--branch edge channels
forming the line junction have opposite chirality and belong to QH
systems with fractional filling factors $1/(m_1+1)$ and $1/(m_2+1)$ with
even integers $m_1\ne m_2$. We assume the junction region to be clean
and of finite length $L$, having edge--channel leads attached that
contact to four reservoirs where transport measurements can be
performed. This sample geometry is experimentally realizable in recently
grown corner junctions between mutually orthogonal 2D electron
systems~\cite{mattcorner1,mattcorner2}. To enable tunneling transport
between them, the two edge channels have to be close enough in space,
which typically facilitates strong interchannel Coulomb interactions
within the junction region. This turns out to significantly affect the
junction conductance when the effective filling factor $\tilde\nu=1/|m_1
-m_2|$ is equal to $1/2$ or $1/4$. Using bosonization and
refermionization techniques, we succeed for both cases in mapping the
originally strongly interacting line--junction system onto a system of
noninteracting fermions. For $\tilde\nu=1/2$, these new quasiparticles
are fictitious chiral spin--$1/2$ fermions which have no direct physical
meaning. Only observables related to their pseudo--spin degree of
freedom correspond to measurable quantities. In the other case of
$\tilde\nu=1/4$, two Majorana fermions having different velocities $v_+$
and $v_-$ turn out to be the fundamental particle excitations in the
line junction. The existence of a velocity splitting directly results in
oscillations of the differential junction conductance as a function of
the transport voltage, which can be detected experimentally. Its
observation would confirm the existence of yet another type of exotic
quasiparticle in low--dimensional systems.

This article is organized as follows. We start, in Sec.~\ref{modelsec},
by specifying the model for an isolated QH line junction and apply the
techniques of bosonization and refermionization to obtain its solution.
The emergence of new quasiparticles will be elucidated. In the following
Sec.~\ref{leadsec}, the coupling of such a line junction to external
edge--channel leads is considered. We derive relations between the
chemical potentials in the leads to quantities describing the line
junction that take full account of charging effects. These results are
then applied in Sec.~\ref{transpsec} to calculate transport properties
of line junctions, with particular focus on the cases $\tilde\nu=1/2$
and $\tilde\nu=1/4$. A summary and conclusions are presented in
Sec.~\ref{conclude}. This article provides full details of and extends
results reported in a previous short publication~\cite{uz:prl:03}.

\section{Effective low--energy model for a quantum--Hall line junction}
\label{modelsec}

A single branch of low--lying edge excitations exists in QH samples at
the Laughlin series of filling factors $\nu_m=1/(m+1)$. These form
chiral 1D electron systems that can be described~\cite{wen:int:92},
using the bosonization~\cite{stone:bos} approach, by a single chiral
boson field $\phi_m(x)$. The (suitably normal--ordered) electronic
charge density at a location $x$ along the edge is given by $\varrho_m
(x)=\,\,:\!\!\psi_m^\dagger(x)\psi_m(x)\!\!:\,\,=\sqrt{\nu_m}\,
\partial_x\phi_m(x)/(2\pi)$. Its dynamics is determined by the
Hamiltonian
\begin{equation}\label{kinhamilt}
H_m=\frac{\hbar v_m}{4\pi}\,\int d x\,\, \left(\partial_x\phi_m\right)^2
\quad .
\end{equation}
The edge velocity $v_m$ is the sum of a one--electron contribution
$v_{\text{F}}$, which is proportional to the slope of the external
potential confining electrons in the QH sample, and an interaction
contribution $\nu_m U/(2\pi\hbar)$ that typically dominates in the
long--wave--length limit~\cite{uz:prb:96}. The chirality of a QH edge is
manifested by the fact that disturbances in the electronic charge
density propagate only in one particular direction along the sample
perimeter. Mathematically, this is expressed by the canonical
commutation relations $\big[\phi_m(x)\, ,\, \phi_m(x^\prime)\big]=i\chi
\,\pi\,\text{sgn}(x-x^\prime)$ for the boson field $\phi_m$. The
chirality parameter $\chi$ assumes the value $+1$ ($-1$) for
right(left)--movers. An especially useful property of chiral 1D systems
is the complete equivalence of their descriptions in terms of bosonic
and fermionic degrees of freedom~\cite{shankar:pol:95,vondelft}. For
single--branch QH edges, this is expressed by the bosonization
identity\cite{wen:int:92} of the electron annihilation operator,
\begin{equation}\label{singbos}
\psi_m(x) = \sqrt{z_m}\,{\mathcal F}_m \, \exp\left\{i x\,\frac{Y}
{\ell^2} + i\chi \, \frac{\phi_m(x)}{\sqrt{\nu_m}}\right\}\quad .
\end{equation}
Here $z_m$ is a normalization constant, and $Y$ denotes the
guiding--center location of electrons at the Fermi energy. The Klein
factor ${\mathcal F}_m$ acts as a ladder operator for the electron
number and ensures fermionic statistics of bosonized electron operators
from different QH edges or edge branches~\cite{vondelft}.

We are interested in studying a QH line junction that is formed when
two parallel fractional--QH edge channels with opposite chirality are
coupled by uniform tunneling along a finite length $L$. Such a junction
can be realized, e.g., by fabricating two 2D electron systems that are
laterally separated~\cite{kang:nat:99} or form a corner
junction~\cite{mattcorner1,mattcorner2}. In the junction region, the
charge densities and, hence, respective bosonic fields $\phi_{m_1}$ and
$\phi_{m_2}$ describing the two edges are coupled via interactions and
tunneling. The total Hamiltonian of the line junction is thus given by
$H_{\text{J}}=H_{\text{LL}}+H_{\text{tun}}$, where
\begin{subequations}
\begin{eqnarray}
H_{\text{LL}} &=& H_{m_1} + H_{m_2} + H_{\text{int}} \quad , \\
H_{\text{int}}  &=& \frac{\sqrt{\nu_{m_1}\nu_{m_2}}}{4\pi^2}\, U \int_{-
\frac{L}{2}}^{\frac{L}{2}} d x \,\, \partial_x\phi_{m_1}\,\partial_x
\phi_{m_2} \quad ,\\
H_{\text{tun}} &=& \int_{-\frac{L}{2}}^{\frac{L}{2}} dx \,\,\left\{ t\,
\psi_{m_1}^\dagger\psi_{m_2} + \text{H.c.} \right\} \quad .
\end{eqnarray}
\end{subequations}
Here we have assumed equal strengths for intra-- and inter--channel
interactions which is the case for typical line junctions.

The part $H_{\text{LL}}$ of the line--junction Hamiltonian can be
diagonalized in a straightforward manner. For the case $m_1=m_2$, the
familiar phase--field description~\cite{fdmh:jpc:81} of a nonchiral
Luttinger liquid is recovered in the typical situation where the Coulomb
matrix element dominates the bare velocities $v_{\text{F}j}$
($j=1,2$). Addition
of the tunneling term $H_{\text{tun}}$ in its bosonized form yields an
orthodox sine--Gordon model whose properties have been studied
extensively~\cite{gog:bos}. We do not discuss this nonchiral case here
any further. Instead, we focus on situations where $m_1\ne m_2$. Then
the line junction is intrinsically chiral, also in the limit of strong
Coulomb interactions.\footnote{A description in terms of nonchiral phase
fields would only be possible at the special (but unphysical) point in
parameter space where $U=2\pi\hbar(\nu_{m_2}-\nu_{m_1})(v_{\text{F}1}-
v_{\text{F}2})$.} Instead, we can write $H_{\text{LL}}$ as a sum
of independent contributions from two chiral normal modes,
\begin{equation}
H_{\text{LL}} = \frac{\hbar}{4\pi}\int_{-\frac{L}{2}}^{\frac{L}{2}} dx
\,\, \left\{v_{\text{a}}\left(\partial_x\phi_{\text{a}}\right)^2 +
v_{\text{b}}\left(\partial_x\phi_{\text{b}}\right)^2\right\}\quad .
\end{equation}
The normal--mode boson fields $\phi_{\text{a}}$ and $\phi_{\text{b}}$
obey the commutation relations $\big[\phi_{\text{a,b}}(x)\, ,\,
\phi_{\text{a,b}}(x^\prime)\big]=i\chi_{\text{a,b}}\,\pi\,\text{sgn}(x-
x^\prime)$ with $\chi_{\text{a}}=-\chi_{\text{b}}$. Assuming the
realistic limit where the bare edge velocities are much smaller than the
Coulomb matrix element, we find the expressions
\begin{subequations}
\begin{eqnarray}
v_{\text{a}}&=& \frac{|\nu_{m1}-\nu_{m2}| U}{2\pi\hbar}+\frac{\nu_{m_1}
v_{\text{F}1} + \nu_{m_2} v_{\text{F}2}}{|\nu_{m1}-\nu_{m2}|} \quad ,\\
v_{\text{b}}&=& \frac{\nu_{m_2}v_{\text{F}1} + \nu_{m_1} v_{\text{F}2}}
{|\nu_{m1}-\nu_{m2}|} \quad , \\ 
\label{chiralities}
\chi_{\text{b}}&=&-\chi_1\,\text{sgn}(\nu_{m_1}-\nu_{m_2}) \quad ,\\
\phi_{\text{a}}&=&\frac{(\sqrt{\nu_{m_1}}-\varepsilon \sqrt{\nu_{m_2}})
\phi_{m_1} + (\sqrt{\nu_{m_2}}-\varepsilon \sqrt{\nu_{m_1}})\phi_{m_2}}
{\sqrt{|\nu_{m_1}-\nu_{m_2}|}}\; ,\\
\phi_{\text{b}}&=&\frac{(\sqrt{\nu_{m_2}}+\varepsilon \sqrt{\nu_{m_1}})
\phi_{m_1} + (\sqrt{\nu_{m_1}}+\varepsilon \sqrt{\nu_{m_2}})\phi_{m_2}}
{\sqrt{|\nu_{m_1}-\nu_{m_2}|}}\; .
\end{eqnarray}
\end{subequations}
Here we have defined a small parameter (of order $\hbar v_{\text{F}j}/
U$)
\begin{equation}\label{eps-def}
\varepsilon = \frac{\sqrt{\nu_{m_1}\nu_{m_2}}}{|\nu_{m_1}-\nu_{m_2}|}
\frac{v_{\text{F}1}+v_{\text{F}2}}{v_{\text{a}}} \quad 
\end{equation}
and neglected terms of quadratic order in $\varepsilon$.\footnote{As it
would be unphysical to neglect the velocity $v_{\text{b}}$ of the slow
normal mode, all terms to first order in the small quantities have to be
kept for consistency. We thank C.~Chamon for focusing our attention on
this point. Note that, for this reason, the normal modes of
$H_{\text{LL}}$ are only approximately, and not exactly, equal to the
familiar charged and neutral modes defined, e.g., in
Ref.~[\onlinecite{uz:prl:03}].}

After bosonization, the tunneling Hamiltonian reads
\begin{equation}
H_{\text{tun}}= 2|t|\sqrt{z_{m_1}z_{m_2}}\int_{-\frac{L}{2}}^{\frac{L}
{2}} dx \,\, \cos\left(\frac{\phi_{\text{n}}}{\sqrt{\tilde\nu}}+x\frac
{\Delta}{\ell^2}\right)\quad ,
\end{equation}
where we have absorbed the phase of the tunneling matrix element into
the neutral--mode bosonic field
\begin{equation}
\phi_{\text{n}}=\frac{\sqrt{\nu_{m_2}}\,\phi_{m_1} + \sqrt{\nu_{m_1}}\,
\phi_{m_2}}{\sqrt{|\nu_{m_1}-\nu_{m_2}|}}-\chi_1\sqrt{\tilde\nu}\,
{\mathrm{arg}}(t)\quad .
\end{equation}
The abbreviation $\tilde\nu=\nu_{m_1}\nu_{m_2}/|\nu_{m_1}-\nu_{m_2}|
\equiv 1/|m_1-m_2|$ has the meaning of an effective junction filling
factor, and $\Delta=Y_1-Y_2$ is a measure for the width of the line
junction. To first order in small quantities defined above, the neutral
mode is given in terms of the normal modes of $H_{\text{LL}}$ as
$\phi_{\text{n}}=\phi_{\text{b}} - \varepsilon \phi_{\text{a}}$.
(Here we have used the freedom to redefine the normal--mode fields by a
constant shift to absorb the phase of the tunneling matrix element in
$\phi_{\text{b}}$.) If we furthermore assume weak enough tunneling such
that $t\ell\ll U$, then terms in $H_{\text{tun}}$ that couple the normal
modes of $H_{\text{LL}}$ are of second order in small quantities. As we
will see below, this is generally the relevant physical situation and
even required in the case $\tilde\nu=1/4$. In what follows we restrict
ourselves to the approximation $\varepsilon\rightarrow 0$ (corresponding
to $U\rightarrow \infty$). We note, however, that corrections to leading
order in $\varepsilon$ can be included perturbatively, yielding small
corrections to our final results which do not affect them significantly.
The Hamiltonian of the line junction is then diagonalized, to a good
approximation, by the normal modes $\phi_{\text{a}}$ and
$\phi_{\text{b}}$ of $H_{\text{LL}}$, and we find $H_{\text{J}}=
H_{\text{a}}+H_{\text{b}}^{\left(\frac{1}{\tilde\nu}\right)}$, where
\begin{subequations}
\begin{equation}
H_{\text{a}} = \frac{\hbar v_{\text{a}}}{4\pi}\int_{-\frac{L}{2}}^{\frac
{L}{2}} dx \,\, \left(\partial_x\phi_{\text{a}}\right)^2 \quad ,
\end{equation}
\begin{widetext}
\begin{equation}
H_{\text{b}}^{\left(\frac{1}{\tilde\nu}\right)} = \int_{-\frac{L}
{2}}^{\frac{L}{2}} dx \,\,\left\{\frac{\hbar v_{\text{b}}}{4\pi}\left(
\partial_x\phi_{\text{b}}\right)^2 + 2|t|\sqrt{z_{m_1}z_{m_2}}\cos\left(
\frac{\phi_{\text{b}}}{\sqrt{\tilde\nu}}+x \frac{\Delta}{\ell^2}\right)
\right\} \quad .
\end{equation}
\end{widetext}
\end{subequations}
The fast normal mode, labeled $\text{a}$, turns out to be free and
unaffected by tunneling. This is quite clear physically, as this mode is
closely related to the total charge density in the line junction
($\rho_{\text{a}}\equiv\frac{\partial_x
\phi_{\text{a}}}{2\pi}$), which
is left invariant in any tunneling process. As we are only interested in
studying tunneling transport, we do not consider the fast mode any
further. The slow mode, being approximately equal to the neutral mode
which measures the {\em difference\/} in electron densities for the two
edge branches forming the junction, has a dynamics that is strongly
influenced by the tunneling term. Its Hamiltonian $H_{\text{b}}^{\left(
\frac{1}{\tilde\nu}\right)}$ is that of a chiral sine--Gordon model
which has been studied in different contexts before~\cite{uz:prl:99,
naud:nucl:00}. Such theories can be classified according to the
different values of the effective junction filling factor $\tilde\nu$
that is given here by the inverse of even integers. For $1/\tilde\nu>4$,
the cosine term has been shown\cite{naud:nucl:00} to be irrelevant in a
renormalization--group sense. This means that it does not alter the
excitation spectrum of the free chiral boson theory in any important way
and can therefore be treated as a perturbation. The situations when $1/
\tilde\nu=4$ and $2$, however, turn out to be different. While no
perturbative approach is permissible for these cases, the effect of the
cosine term can nevertheless be calculated, even exactly, using
bosonization identities of the kind expressed in Eq.~(\ref{singbos}).
Namely, it is possible\cite{uz:prl:99,naud:nucl:00} to map the rather
complicated chiral bosonic field theory for the slow mode onto that of
noninteracting fictitious fermions. We proceed to show this in the
following two subsections, as the refermionized description of the
slow--mode dynamics forms the basis for our subsequent transport
calculations.

\subsection{Case $\mathbf{\tilde\nu=1/2}$: Fictitious chiral fermion
tunneling}\label{ficttun}

To solve the chiral sine--Gordon Hamiltonian with $\tilde\nu=1/2$, we
introduce --- for purely mathematical reasons --- a ghost field $\eta(x)
$ that has the same chirality and dynamics as the slow mode
$\phi_{\text{b}}$. It is then possible to define a pair of fictitious
chiral fermions, distinguished by a pseudospin degree of freedom $\sigma
=\uparrow,\downarrow$ using the bosonization identity
\begin{equation}
\Psi_\sigma(x)=\sqrt{z_{\text{b}}}\,{\mathcal F}_\sigma\,\exp\left\{i
\chi_{\text{b}}\,\frac{\eta(x)+\sigma\phi_{\text{b}}(x)}{\sqrt{2}}\right
\}\quad .
\end{equation}
In this new notation, the Hamiltonian $H_{\text{b}}^{(2)}$ of the slow
mode represents tunneling between the two flavors of fictitious
fermions:
\begin{widetext}
\begin{equation}\label{chirtunn}
H_{\text{b}}^{(2)}=\int_{-\frac{L}{2}}^{\frac{L}{2}} dx\,\,\left\{\hbar
v_{\text{b}}\sum_\sigma\Psi_\sigma^\dagger(-i\chi_{\text{b}}\partial_x)
\Psi_\sigma+\tilde t_{m_1m_2}\left[\Psi_\uparrow^\dagger\Psi_\downarrow
e^{-i\chi_{\text{b}}x\Delta/\ell^2} + {\mathrm H.c.}\right]\right\}
\quad .
\end{equation}
\end{widetext}
The tunneling strength $\tilde t_{m_1m_2}=|t|\sqrt{z_{m_1}z_{m_2}}/
z_{\text{b}}$ is generally different, because of
chiral--Luttinger--liquid properties, from the matrix element for
tunneling between the original line--junction edge
channels.\footnote{This is due to the power--law system--size dependence
of the normalization factors $z_{m_j}$. See, e.g., J.~M. Kinaret, Y.
Meir, N.~S. Wingreen, P.~A. Lee, and X.~G. Wen, Phys. Rev. B {\bf 46},
4681 (1992).} Physical observables are expressed in terms of
pseudospin--related quantities, while the charge degree of freedom for
fictitious fermions remains hidden from measurement. For example, the
spin density and the density associated with the slow mode are
equivalent:
\begin{equation}\label{fictDiracdens}
\Psi_\uparrow^\dagger\Psi_\uparrow-\Psi_\downarrow^\dagger
\Psi_\downarrow=\sqrt{2}\varrho_{\text{b}}\equiv\frac{\partial_x
\phi_{\text{b}}}{\sqrt{2}\pi}\quad .
\end{equation}
The representation of $H_{\text{b}}^{(2)}$ in terms of the fictitious
chiral pseudo--spin--1/2 fermion, being quadratic in this field, makes
it possible to treat transport straightforwardly. This will be discussed
below in Section~\ref{transpsec1}.

\subsection{Case $\mathbf{\tilde\nu=1/4}$: Velocity--split Majorana
fermions}\label{majferm}

A bosonization identity of the type given in Eq.~(\ref{singbos}) can be
used to define a Dirac fermion in terms of the chiral boson field
$\phi_{\text{b}}$:
\begin{equation}
\psi_{\text{b}}(x)=\sqrt{z_{\text{b}}}\,{\mathcal F}_{\text{b}}\,
\exp\left\{i\chi_{\text{b}}\left[x\,\frac{\Delta}{2\ell^2}+
\phi_{\text{b}}(x)\right]\right\} \quad .
\end{equation}
The density $\varrho_{\text{b}}$ of the slow mode is related to the
normal--ordered density $:\psi_{\text{b}}^\dagger\psi_{\text{b}}\!:$ of
the new fictitious fermion via
\begin{equation}
\varrho_{\text{b}}\equiv\frac{\partial_x\phi_{\text{b}}}{2\pi}=\,
:\psi_{\text{b}}^\dagger\psi_{\text{b}}\!:-\frac{\Delta}{2\ell^2}\quad .
\end{equation}
With the help of the relation\cite{naud:nucl:00}
\begin{equation}
\psi_{\text{b}}\, i\partial_x\,\psi_{\text{b}} = -2\pi\chi_{\text{b}}\,
z_{\text{b}}^2\,{\mathcal F}_{\text{b}}^2\,\exp\left\{i\chi_{\text{b}}
\left[2 \phi_{\text{b}} + x\,\frac{\Delta}{\ell^2}\right]\right\} ,
\end{equation}
the Hamiltonian of the slow mode can be rewritten as
\begin{widetext}
\begin{equation}\label{pwavesup}
H_{\text{b}}^{(4)}=\int_{-\frac{L}{2}}^{\frac{L}{2}} dx\,\,\left\{\hbar
v_{\text{b}}\psi_{\text{b}}^\dagger (-i\chi_b\partial_x)\psi_{\text{b}}
 - \hbar v_{\text{b}}\,\frac{\Delta}{2\ell^2}\,\psi_{\text{b}}^\dagger
\psi_{\text{b}} + \frac{\hbar v_{\text{t}}}{2}\left[ \psi_{\text{b}}(-i
\chi_{\text{b}} \partial_x ) \psi_{\text{b}}+\psi_{\text{b}}^\dagger(-i
\chi_{\text{b}}\partial_x)\psi_{\text{b}}^\dagger\right]\right\}\quad ,
\end{equation}
\end{widetext}
where the tunneling matrix element has been absorbed into the velocity
parameter $v_{\text{t}}=|t|\sqrt{z_{m_1}z_{m_2}}/\pi\hbar z_{\text{b}}^2
$. In the representation of the fictitious fermion $\psi_{\text{b}}$,
$H_{\text{b}}^{(4)}$ looks like the Bogoliubov--de~Gennes
Hamiltonian~\cite{degennes} of a spinless p--wave superconductor. It
differs from similar systems considered previously~\cite{nick:prb:00,
stoneprep} by its chiral 1D nature. We do not explicitly pursue the
superconducting analogy here any further (although the formalism
employed below could be phrased within such a framework). Instead, we
use the fact that the real and imaginary parts of a Dirac fermion are
Majorana fermions and define fields $\xi_\pm\equiv\xi_\pm^\dagger$ via
\begin{subequations}
\begin{eqnarray}
\xi_+&=&\frac{\psi_{\text{b}}+\psi_{\text{b}}^\dagger}{\sqrt{2}}\quad ,
\\ \xi_-&=& \frac{\psi_{\text{b}}-\psi_{\text{b}}^\dagger}{\sqrt{2}\,i}
\quad .
\end{eqnarray}
\end{subequations}
Note that $:\!\psi_{\text{b}}^\dagger\psi_{\text{b}}\!: = i\xi_+\xi_- $.
The expression for the Hamiltonian of the slow mode reads
\begin{equation}
H_{\text{b}}^{(4)}=\int_{-\frac{L}{2}}^{\frac{L}{2}} dx\,\,\left\{
\frac{\hbar}{2}\sum_{r=\pm} v_r\,\xi_r(-i\chi_{\text{b}}\partial_x)\xi_r
-i\,\frac{\hbar v_{\text{b}}\Delta}{\ell^2}\, \xi_+ \,\xi_- \right\}\, ,
\end{equation}
with different velocities $v_r=v_{\text{b}}+rv_{\text{t}}$ for the two
Majorana fields. Hence, while the slow mode is characterized by a single
velocity $v_{\text{b}}$, tunneling leads to the generation of two
different velocities for the resulting quasiparticles that turn out to
be Majorana fermions. A dynamically generated velocity splitting of this
type has been found before in tunnel--coupled interacting quantum
wires~\cite{fink:prb:93,fab:prl:93,uz:prb:02a} and quantum--Hall
bilayers~\cite{naud:nucl:00}. It is reminiscent of the spin--charge
separation expected to occur in interacting 1D electron
systems~\cite{voit:reprog:94}. The representation of $H_{\text{b}}^{(4)}
$ in terms of the fictitious noninteracting Majorana fermions enables
treatment of transport through the line junction, which will be shown
below in Sec.~\ref{transpsec2}. Diagonalization of $H_{\text{b}}^{(4)}$
is straightforward. Its spectrum has two branches,
\begin{equation}
E_{k,\pm}=\hbar\chi_{\text{b}} k \left( v_{\text{b}} \pm v_{\text{t}}
\sqrt{1+r_k^2}\right)\, ,
\end{equation}
with $r_k=v_{\text{b}}\Delta/(v_{\text{t}} k \ell^2)$. The corresponding
eigenstates are given, using a spinor notation in the basis of Majorana
fermions $\xi_\pm$, as
\begin{subequations}
\begin{eqnarray}
\left(\begin{array}{c} \xi_+ \\ \xi_- \end{array}\right)_+ &=& \left(
\begin{array}{c} c_k \\ i \chi_{\text{b}} s_k \end{array} \right)
e^{i k x} \, , \\ \left(\begin{array}{c} \xi_+ \\ \xi_- \end{array}
\right)_- &=& \left(\begin{array}{c} i \chi_{\text{b}} s_k \\ c_k 
\end{array} \right) e^{i k x} \, .
\end{eqnarray}
\end{subequations}
Here we used the abbreviations
\begin{equation}
\left. \begin{array}{c} c_k \\ s_k \end{array}\right\} =
\frac{\sqrt{\sqrt{1+r_k^2}\pm 1}}{\sqrt{2\sqrt{1+r_k^2}}}\quad .
\end{equation}

\section{Tunneling current, chemical potential, and coupling to
voltages} \label{leadsec}

To be able to calculate conductances, it is necessary to treat the
nonequilibrium situation where a finite voltage is applied to the line
junction. This requires proper definition of operators for currents and
chemical potentials within the junction region, which we set out to do
in the first part of this section. Our results are then applied to
relate these quantities to externally adjustable lead voltages.

A standard calculation yields the expression for the tunneling current
flowing from edge channel 1 to edge channel 2 within the line junction
as
\begin{equation}
I_{\text{J}}=\frac{i}{\hbar} \int_{-\frac{L}{2}}^{\frac{L}{2}} dx \,\,
\left\{t\, \psi_{m_1}^\dagger\psi_{m_2} - \text{H.c.} \right\} \quad .
\end{equation}
After bosonization, and within the same approximations used above, i.e.,
for small $\varepsilon$ [see Eq.~(\ref{eps-def})], we find the following
expression for the transport current through the line junction:
\begin{equation}
I_{\text{J}}^{\left(\frac{1}{\tilde\nu}\right)}=\chi_1\frac{2|t|}{\hbar}
\,\sqrt{z_{m_1}z_{m_2}}\int_{-\frac{L}{2}}^{\frac{L}{2}}\!\! dx\,\,\sin
\left(\frac{\phi_{\text{b}}}{\sqrt{\tilde\nu}}+x\frac{\Delta}{\ell^2}
\right)\, .
\end{equation}
It comes as no surprise that the current, as the tunneling Hamiltonian,
depends only on the slow normal mode. Hence, only the latter features in
our transport calculation. Defining the spatially varying partial
current
\begin{widetext}
\begin{equation}
I_{\text{b}}^{\left(\frac{1}{\tilde\nu}\right)}(x)=\chi_1\frac{|t|}
{\hbar\sqrt{\tilde\nu}}\, \sqrt{z_{m_1}z_{m_2}}\,\int_{-\frac{L}
{2}}^{\frac{L}{2}} dx^\prime\,\,\sin\left(\frac{\phi_{\text{b}}(
x^\prime)}{\sqrt{\tilde\nu}}+x^\prime\frac{\Delta}{\ell^2}\right)\,
{\mathrm{sgn}}(x-x^\prime)
\end{equation}
\end{widetext}
turns out to be useful for later.
Obviously, $I_{\text{b}}^{\left(\frac{1}{\tilde\nu}\right)}(L/2)=-
I_{\text{b}}^{\left(\frac{1}{\tilde\nu}\right)}(-L/2)\equiv
I_{\text{J}}^{\left(\frac{1}{\tilde\nu}\right)}/(2\sqrt{\tilde\nu})$.

The local chemical potential of the slow normal mode can be defined in
the usual way~\cite{safi:epjb:99} as the functional derivative of the
Hamiltonian with respect to density: $\mu_{\text{b}}^{\text{(J)}}(x)=
\delta H_{\text{J}}/\delta\varrho_{\text{b}}(x)$. Using the commutation
relations for chiral boson fields, it is straightforward to prove that
$\chi_{\text{b}}\phi_{\text{b}}$ is canonically conjugate to
$\varrho_{\text{b}}$, and therefore
\begin{equation}
\frac{\delta\mathcal O}{\delta\rho_{\text{b}}(x)}=i\chi_{\text{b}}\,
\big[\phi_{\text{b}}(x)\, , \, {\mathcal O}\big]
\end{equation}
for an arbitrary functional $\mathcal O$. Application of this identity
yields an expression for the local chemical potential,
\begin{equation}\label{chempot}
\mu_{\text{b}}^{\text{(J)}}(x)=2\pi\hbar\left\{v_{\text{b}}
\varrho_{\text{b}}(x)-\chi_1\, I_{\text{b}}^{\left(\frac{1}{\tilde\nu}
\right)}(x)\right\}\quad .
\end{equation}
As a special case of this equation, we find the chemical potentials at
the endpoints of the junction to be
\begin{equation}\label{endpoints}
\mu_{\text{b}}^{\text{(J)}}\left(\pm\frac{L}{2}\right)=2\pi\hbar\left\{
v_{\text{b}}\varrho_{\text{b}}\left(\pm\frac{L}{2}\right)\mp\frac{\chi_1
\, I_{\text{J}}^{\left(\frac{1}{\tilde\nu}\right)}}{2\sqrt{\tilde\nu}}
\right\}\quad .
\end{equation}

The continuity equation for the slow--mode density can be derived in a
similar way,
\begin{subequations}
\begin{eqnarray}
\frac{d}{d\tau}\,\varrho_{\text{b}}&=&\partial_\tau\varrho_{\text{b}}+
\frac{i}{\hbar}\big[H_{\text{J}}\, ,\, \varrho_{\text{b}}\big]\quad , \\
&=& \partial_\tau\varrho_{\text{b}}+\chi_b\partial_x\left\{v_{\text{b}}
\varrho_{\text{b}}-\chi_1\, I_{\text{b}}^{\left(\frac{1}{\tilde\nu}
\right)}\right\}\quad . \label{contequ}
\end{eqnarray}
\end{subequations}
In the stationary limit where $\partial_\tau\varrho_{\text{b}}=0$,
Eqs.~(\ref{chempot}) and (\ref{contequ}) imply the exact relation
\begin{equation}
\partial_x\mu_{\text{b}}^{\text{(J)}}=2\pi\hbar\chi_{\text{b}}\,\frac{d}
{d\tau}\,\varrho_{\text{b}}\quad .
\end{equation}
Integrating it, and observing the relation $\sqrt{\tilde\nu}\frac{d}
{d\tau}\int_x\varrho_{\text{b}}=I_{\text{J}}^{\left(\frac{1}{\tilde\nu}
\right)}{\mathrm{sgn}}(\nu_{m_1}-\nu_{m_2})$, we find
\begin{equation}
\mu_{\text{b}}^{\text{(J)}}\left(\frac{L}{2}\right)-
\mu_{\text{b}}^{\text{(J)}}\left(-\frac{L}{2}\right)=-2\pi\hbar\chi_1\,
\frac{I_{\text{J}}^{\left(\frac{1}{\tilde\nu}\right)}}{\sqrt{\tilde\nu}}
\quad ,
\end{equation}
and comparison with Eqs.~(\ref{endpoints}) yields
\begin{equation}\label{statdens}
\varrho_{\text{b}}\left(\frac{L}{2}\right)=\varrho_{\text{b}}\left(-
\frac{L}{2}\right)\equiv \bar\varrho_{\text{b}}\quad .
\end{equation}
The periodic boundary condition expressed in Eq.~(\ref{statdens}) is a
nontrivial property of the stationary state that enables us to treat the
line junction as separate from any edge--channel leads that couple it to
external reservoirs. The effect of external voltages will be to set the
appropriate value of $\bar\varrho_{\text{b}}$ consistent with the value
$I_{\text{J}}$ for the line--junction current. A derivation of these
relations will be presented in the following paragraphs.

\begin{figure}[b]
\includegraphics[width=2.7in]{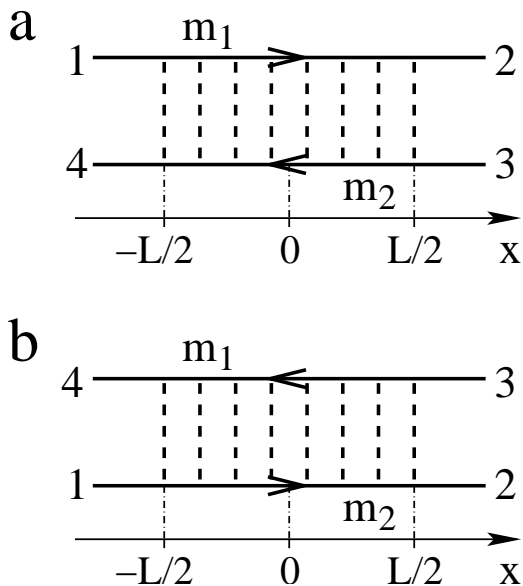}
\caption{Line junction attached to chiral edge--channel leads. The
chemical potentials of incoming lead branches (1 and 3) are fixed
experimentally by external reservoirs. These, together with the
tunneling current in the junction region (defined to flow from the $m_1$
branch to the $m_2$ branch and indicated by vertical dashed lines),
determine the boundary condition for the slow normal--mode density at
the junction boundaries $x=\pm L/2$. Panels a and b show the two
possible chiralities of the junction that can be distinguished, e.g., by
the chirality of the $m_1$ branch: $\chi_1=+1$ ($-1$) for a (b).
\label{sketchprb}}
\end{figure}

Edge channels forming the line junction typically exist beyond the
junction region but are not coupled anymore via tunneling or
interactions. These incoming and outgoing edge branches serve as
noninteracting leads that couple the junction to external reservoirs
with experimentally controllable chemical potentials. (See the figure in
Ref.~[\onlinecite{uz:prl:03}] and Fig.~\ref{sketchprb} below for
illustration.) The Hamiltonian for the lead regions $|x|>L/2$ is
therefore given by that of two uncoupled edges; $H_{\text{E}}=H_{m_1}+
H_{m_2}$. As the separate chiral edge densities for each edge branch are
just linear combinations of the densities $\varrho_{\text{a}}$ and
$\varrho_{\text{b}}$ that are the normal modes in the junction, we can
write down an expression for $H_{\text{E}}$ in terms of the latter. In
the limit of strong intra--edge Coulomb interactions, which is the
physically relevant situation that we have been considering all along,
it reads
\begin{widetext}
\begin{equation}
H_{\text{E}}=\frac{U}{2|\nu_{m_1}-\nu_{m_2}|}\int_{|x|>\frac{L}{2}} dx
\,\, \left\{(\nu_{m_1}^2+\nu_{m_2}^2)\varrho_{\text{a}}^2 + 2 \nu_{m_1}
\nu_{m_2}\varrho_{\text{b}}^2 - 2 \sqrt{\nu_{m_1}\nu_{m_2}}(\nu_{m_1}+
\nu_{m_2})\varrho_{\text{a}}\varrho_{\text{b}}\right\}\quad .
\end{equation}
\end{widetext}
While the a and b modes are certainly not normal modes in the lead
region, it is nevertheless possible to define their chemical potentials
via $\mu_{\text{a,b}}^{\text{(E)}}(x)=\delta H_{\text{E}}/\delta
\varrho_{\text{a,b}}(x)$. As $H_{\text{E}}$ is a nondiagonal quadratic
form when expressed in terms of $\varrho_{\text{a}}$ and
$\varrho_{\text{b}}$, each of the chemical potentials
$\mu_{\text{a}}^{\text{(E)}}$ and $\mu_{\text{b}}^{\text{(E)}}$ depends
on both densities. In other words, there exists a finite
cross--capacitance between the a and b modes in the lead region.
Re--expressing $\mu_{\text{a,b}}$ in terms of the original chiral edge
densities, we find
\begin{subequations}\label{chemrel}
\begin{eqnarray}
\frac{\mu_{\text{a}}^{\text{(E)}}}{U} &=& \frac{\nu_{m_1}\varrho_{m_1}-
\nu_{m_2}\varrho_{m_2}}{\sqrt{|\nu_{m_1}-\nu_{m_2}|}}\,\,{\mathrm{sgn}}
(\nu_{m_1}-\nu_{m_2}) \,\, , \\
\frac{\mu_{\text{b}}^{\text{(E)}}}{U} &=& -\sqrt{\tilde\nu}\left(
\varrho_{m_1}-\varrho_{m_2}\right)\,{\mathrm{sgn}}(\nu_{m_1}-\nu_{m_2})
\,\, ,
\end{eqnarray}
\end{subequations}
where again terms of order $\varepsilon$ have been neglected. The
expressions given in Eq.~(\ref{chemrel}) are useful because the local
chemical potential in each of the chiral edge branches within the lead
region is directly related to the local chiral edge density in that
branch via $\mu_j=U\,\varrho_j$, within the limit of strong intra--edge
Coulomb interactions considered here. Experimentally, two of the four
lead branches' chemical potentials are controlled by attachment to
external reservoirs, namely those having chiral edge--density waves
propagating {\em away\/} from a reservoir. We adopt the convention that
the incoming (outgoing) edge lead for $x<-L/2$ is labeled 1 (4),
whereas the incoming (outgoing) edge lead for $x>L/2$ is labeled 3 (2).
(See Fig.~\ref{sketchprb}.) We find then from Eqs.~(\ref{chemrel})
\begin{equation}
\left.\mu_{\text{b}}^{\text{(E)}}\right|_{x\stackrel{<}{>}\mp\frac{L}{2}
} = \chi_{\text{b}}\,\sqrt{\tilde\nu}\left(\mu_{1/2}-\mu_{4/3}\right)
\quad .
\end{equation}
Requiring continuity of the chemical potentials $\mu_{\text{a,b}}$ at
the lead--junction interfaces, and using Eqs.~(\ref{endpoints}),
(\ref{statdens}) and (\ref{chiralities})
as well as the constancy of $\mu_{\text{a}}$ throughout
the line junction, we find the relations
\begin{subequations}
\begin{eqnarray}
4\pi\hbar\chi_{\text{b}}\, v_{\text{b}}\bar\varrho_{\text{b}}&=&
\sqrt{\tilde\nu}(\mu_1-\mu_3 +\mu_2 - \mu_4)\, , \\
\frac{2\pi\hbar\, I_{\text{J}}^{\left(\frac{1}{\tilde\nu}\right)}}
{{\mathrm{sgn}}(\nu_{m_1}-\nu_{m_2})}&=&-\tilde\nu(\mu_1+\mu_3-\mu_2-
\mu_4)\, , \\ \label{chirvar}
\nu_{m_1}(\mu_{1(4)}-\mu_{2(3)})&=&\nu_{m_2}(\mu_{4(1)}-\mu_{3(2)})\, .
\end{eqnarray}
\end{subequations}
Here we used a compact notation in Eq.~(\ref{chirvar}) to show the cases
corresponding to both line--junction chiralities $\chi_1=+(-)1$ in one
line. Straightforward elimination yields an equation for the
voltage drop across the junction,
\begin{equation}\label{fundament}
\frac{\mu_1-\mu_3}{2\pi\hbar} = \frac{\chi_{\text{b}}}{\sqrt{\tilde\nu}}
\,\, v_{\text{b}}\bar\varrho_{\text{b}} + \chi_1\,\frac{\nu_{m_1}+
\nu_{m_2}}{2\nu_{m_1}\nu_{m_2}}\,\,I_{\text{J}}^{\left(\frac{1}{\tilde
\nu}\right)}\quad ,
\end{equation}
providing a link between the externally applied voltage $e V=\mu_1-\mu_3
$ and intrinsic line--junction quantities, such as the tunneling current
and the boundary value $\bar\varrho_{\text{b}}$ of the slow--mode
density. This is one of the central results of our work presented here.
In deriving Eq.~(\ref{fundament}), we have fully taken into account
charging of the line junction arising from its coupling to leads.
Solving it together with the intrinsic dynamics of the slow mode
(induced by its Hamiltonian $H_{\text{b}}^{\left(\frac{1}{\tilde\nu}
\right)}$) completely determines the current.

\section{Results for transport through line junctions}
\label{transpsec}

We have seen, in Sec.~\ref{modelsec}, that the tunneling current through
the line junction is determined by the dynamics of the slow mode only.
Relating its Hamiltonian to that of free quasiparticles, which are Dirac
fermions for $\tilde\nu=1/2$ and Majorana fermions for $\tilde\nu=1/4$,
we are able to calculate its dynamics exactly. With the help of
Eq.~(\ref{fundament}), quantities determining this dynamics are related
to the externally applied voltage. Pulling both understandings together,
we find the line--junction conductance. As it is quite different for the
two cases mentioned above, we discuss them in separate subsections.

\subsection{Case $\mathbf{\tilde\nu=1/2}$: Conductance oscillations as a
function of junction length}\label{transpsec1}

Within the refermionization procedure applied in Sec.~\ref{ficttun}, we
find the current through the junction as the tunneling current between
the two flavors of a fictitious chiral Dirac pseudo--spin 1/2 fermion:
\begin{equation}
I_{\text{J}}^{(2)} = -i\chi_{\text{b}}\chi_1\,\frac{\tilde t_{m_1m_2}}
{\hbar}\int_{-\frac{L}{2}}^{\frac{L}{2}} d x\,\,\left\{
\Psi_\uparrow^\dagger\Psi_\downarrow e^{-i\chi_{\text{b}}x\Delta/\ell^2}
 - {\mathrm H.c.}\right\}\, .
\end{equation}
The chemical--potential difference between the $\uparrow$ and
$\downarrow$ branches is related to that of the slow mode via
\begin{equation}
\mu_\uparrow-\mu_\downarrow=\sqrt{2}\,\mu_{\text{b}}^{\text{(J)}}\quad .
\end{equation}
Equation~(\ref{statdens}), together with Eq.~(\ref{fictDiracdens}),
implies a constant density difference  between the two fermion flavors
along the junction. This provides the driving force for the current.

The chiral--tunneling problem can be solved exactly using standard
methods~\cite{dircoupl1,pank:jpc:00,uz:prb:01}. Here we employ the
scattering approach to transport~\cite{landauer2,butt:ibm:88}. Using an
obvious spinor notation for the pseudospin--1/2 fermion wave functions,
the following {\it Ans\"atze} for eigenstates of the Hamiltonian
$H_{\text{b}}^{(2)}$ [displayed in Eq.~(\ref{chirtunn})] in the three
regions $\chi_{\text{b}} x<-L/2$, $\chi_{\text{b}} x> L/2$, and $|x|< 
L/2$ can be written down:
\begin{subequations}
\begin{eqnarray}
&&\left.\left(\begin{array}{c} \Psi_\uparrow \\ \Psi_\downarrow
\end{array}\right)\right|_{\chi_{\text{b}} x< -\frac{L}{2}}=\left(
\begin{array}{c} 1 \\ 0 \end{array}\right) e^{i x \frac{\chi_{\text{b}}
E}{\hbar v_{\text{b}}}}\quad , \\
&&\left.\left(\begin{array}{c} \Psi_\uparrow \\ \Psi_\downarrow
\end{array}\right)\right|_{\chi_{\text{b}} x> \frac{L}{2}}=\left(
\begin{array}{c} t_\uparrow \\ 0 \end{array}\right) e^{i x \frac
{\chi_{\text{b}} E}{\hbar v_{\text{b}}}}+\left(\begin{array}{c} 0 \\
t_\downarrow \end{array}\right) e^{i x \frac{\chi_{\text{b}} E}{\hbar
v_{\text{b}}}} \, , \\
&&\left.\left(\begin{array}{c} \Psi_\uparrow \\ \Psi_\downarrow
\end{array}\right)\right|_{|x|<\frac{L}{2}}=a\left(\begin{array}{c} d_+
\\ d_- \end{array}\right)e^{i x \kappa_+} + b \left(\begin{array}{c}
- d_-^\ast \\ d_+^\ast\end{array}\right) e^{i x \kappa_-} \, ,
\end{eqnarray}
with the abbreviations
\begin{eqnarray}
d_\pm &=& \pm\sqrt{\sqrt{1+\zeta^2}\pm\zeta}\,\, e^{\mp i\chi_{\text{b}}
x \frac{\Delta}{2\ell^2}} \quad , \\
\kappa_\pm &=& \chi_{\text{b}} \, \frac{E\pm \tilde t_{m_1m_2} \sqrt{1
+ \zeta^2}}{\hbar v_{\text{b}}} \quad .
\end{eqnarray}
\end{subequations}
The parameter $\zeta=\hbar v_{\text{b}}\Delta/(2 \tilde t_{m_1m_2}\ell^2
)$ measures the deviation from perfect energy and momentum conservation
for tunneling pseudofermions. Requiring continuity of the wave function
at $x=\pm L/2$ yields a system of linear equations that we can use to
find expressions for the transmission coefficients $t_\uparrow$ and
$t_\downarrow$. Note that, due to chirality, continuity of the wave
function is sufficient to ensure current conservation. As $t_\uparrow$
and $t_\downarrow$ turn out to be independent of energy $E$, the
tunneling current is proportional to the density difference of the 
pseudospin components.  Its explicit expression is
\begin{subequations}
\begin{equation}
I_{\text{J}}^{(2)}= -\left|t_\downarrow\right|^2 \,\sqrt{2}\,
v_{\text{b}}\bar\varrho_{\text{b}}\, {\mathrm{sgn}}(\nu_{m_1}-\nu_{m_2})
\quad ,
\end{equation}
with the transmission probability found from the above calculation as
\begin{equation}\label{transcoeff}
\left|t_\downarrow\right|^2 \equiv T(L)=\frac{\sin^2\left[\frac{\pi L}
{L_t}\sqrt{1+\zeta^2}\right]}{1 + \zeta^2} \quad .
\end{equation}
\end{subequations}
Here $L_t=\pi\hbar v_{\text{b}}/\tilde t_{m_1m_2}$ is a length scale set
by the effective tunneling strength. From Eq.~(\ref{fundament}), we find
the expression for the current through the line junction as a function
of the externally applied voltage
\begin{equation}\label{fullsol2}
I_{\text{J}}^{(2)}=\frac{\chi_1}{2\pi\hbar}\,\,\frac{T(L)}{1+
\frac{\nu_{m_1}+\nu_{m_2}}{2\nu_{m_1}\nu_{m_2}}\, T(L)}\,\,(\mu_1-\mu_3)
\quad .
\end{equation}
With Eq.~(\ref{fullsol2}), we have the full solution of the transport
problem for line junctions with effective filling factor $\tilde\nu=1/2
$. A linear I--V characteristic is obtained, with a conductance
oscillating as a function of junction length $L$ and magnetic field
(through dependence on the parameter $\zeta$). Its maximum value,
reached for $T(L)\to 1$ and given by
\begin{equation}
G_{\text{J}}^{\text{max}}=\frac{e^2}{2\pi\hbar}\,\,{\mathrm{min}}\left\{
\nu_{m_1},\nu_{m_2}\right\}\quad ,
\end{equation}
is smaller than that of an adiabatic point contact between chiral
fractional edge channels considered in Ref.~[\onlinecite{bih:prb:98}],
as should be expected.

\subsection{Case $\mathbf{\tilde\nu=1/4}$: Conductance oscillations as a
function of transport voltage}\label{transpsec2}

Refermionization yields an expression for the tunneling current in terms
of the fictitious Dirac fermion $\psi_{\text{b}}$,
\begin{equation}
I_{\text{J}}^{(4)} = \chi_1\, \frac{v_{\text{t}}}{2}
\int_{-\frac{L}{2}}^{\frac{L}{2}} dx\,\, \left\{\psi_{\text{b}}^\dagger
\partial_x \psi_{\text{b}}^\dagger - \psi_{\text{b}}\partial_x
\psi_{\text{b}}\right\}\, ,
\end{equation}
or the Majorana fermions that diagonalize the slow--mode
Hamiltonian when $\Delta=0$:
\begin{equation}
I_{\text{J}}^{(4)} = \frac{v_{\text{t}}}{2}\int_{-\frac{L}{2}}^{\frac{L}
{2}} dx\,\, \left\{\xi_+ (-i\chi_1\partial_x) \xi_- + \xi_- (-i\chi_1
\partial_x) \xi_+\right\}\, .
\end{equation}
In close analogy with the chiral--tunneling problem encountered for
$\tilde\nu=1/2$ and discussed in Sec.~\ref{transpsec1} above, we solve
the transport problem through the line junction by considering the
scattering of Dirac fermions $\psi_{\text{b}}$ in terms of the Majorana
normal modes. As in the Landauer--B\"uttiker formalism for
transport~\cite{landauer2,butt:ibm:88}, incoming scattering states are
postulated for $\chi_{\text{b}} x<-L/2$, given in terms of a Majorana
spinor notation as
\begin{equation}
\left.\left(\begin{array}{c} \xi_+ \\ \xi_- \end{array}\right)
\right|_{\chi_{\text{b}} x < -\frac{L}{2}} = \left(\begin{array}{c} 1 \\
i \chi_{\text{b}} \end{array} \right) e^{i x (\frac{\chi_{\text{b}} E}
{\hbar v_{\text{b}}}-\frac{\Delta}{\ell^2})} \quad .
\end{equation}
Outgoing scattering states exist for $\chi_{\text{b}} x>L/2$ and are
superpositions of a Dirac--fermion state $\psi_{\text{b}}$ and a Dirac
`hole' state $\psi_{\text{b}}^\dagger$:\footnote{This is not surprising,
as the refermionized slow--mode Hamiltonian $H_{\text{b}}^{(4)}$ in the
representation of Dirac fermions is analogous to the
Bogoliubov--de~Gennes equation for a p--wave superconductor. See
Eq.~(\ref{pwavesup}) and remarks below.}
\begin{widetext}
\begin{equation}
\left.\left(\begin{array}{c} \xi_+ \\ \xi_- \end{array}\right)
\right|_{\chi_{\text{b}} x > \frac{L}{2}} = t_-\left(\begin{array}{c} 1
\\ i \chi_{\text{b}} \end{array} \right)e^{i x (\frac{\chi_{\text{b}} E}
{\hbar v_{\text{b}}}-\frac{\Delta}{\ell^2})} + t_+\left(\begin{array}{c}
1 \\ - i \chi_{\text{b}} \end{array}\right) e^{i x(\frac{\chi_{\text{b}}
E}{\hbar v_{\text{b}}}+\frac{\Delta}{\ell^2})} \quad .
\end{equation}
In the line--junction region, a superposition of the two eigenstates
with energy eigenvalue $E$ is realized:
\begin{equation}
\left.\left(\begin{array}{c} \xi_+ \\ \xi_- \end{array}\right)
\right|_{|x| < \frac{L}{2}} = a \left(\begin{array}{c} c_{k_+} \\i
\chi_{\text{b}} s_{k_+} \end{array} \right) e^{i k_+ x} + b \left(
\begin{array}{c} i \chi_{\text{b}} s_{k_-} \\ c_{k_-} \end{array}
\right) e^{i k_- x} \quad .
\end{equation}
Their respective wave vectors are solutions of $E_{k_\pm,\pm}=E$.
Requiring continuity of the Majorana--spinor wave functions at the
line--junction interfaces $x=\pm L/2$, which simultaneously ensures
current conservation for the chiral--fermion problem considered here,
yields expressions for the transmission amplitudes $t_\pm$. In contrast
to the case $\tilde\nu=1/2$, the latter are energy dependent. The
transport current through the line junction is given by
\begin{equation}\label{currentequ}
I_{\text{J}}^{(4)}=\frac{\chi_1\chi_{\text{b}}}{2\pi\hbar}\int_0^{2\pi
\hbar v_{\text{b}}\bar\varrho_{\text{b}}} d E \,\, \left|t_+(E)\right|^2
\quad ,
\end{equation}
hence we only need the general expression for $|t_+|^2$, which is found
to be
\begin{equation}
|t_+|^2 = 2\frac{\left(\sqrt{1+r_{k_+}^2}-|r_{k_+}|\right)\left(\sqrt{1+
r_{k_-}^2}+|r_{k_-}|\right)}{1+|r_{k_+}| |r_{k_-}| +\sqrt{1+r_{k_+}^2}
\sqrt{1+r_{k_-}^2}}\,\sin^2\left[\frac{1}{2}(k_+ - k_-) L\right] \quad .
\end{equation}
\end{widetext}
Note that the energy dependence of $|t_+|^2$ is implicit through the
dependence of $k_\pm$ on $E$.

\begin{figure}[b]
\includegraphics[width=3.3in]{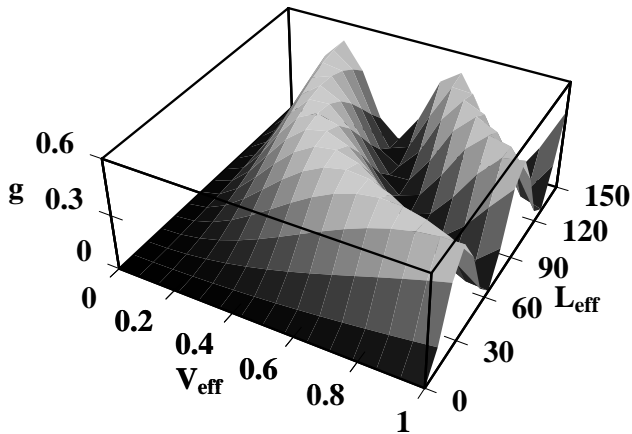}
\caption{Illustration of conductance oscillations. The dimensionless
differential tunneling conductance $g=\frac{2\pi\hbar}{e^2}\frac{dI}{dV}
$ is plotted as a function of $V_{\text{eff}}=\frac{\epsilon\ell}{e}\,
V$ and $L_{\text{eff}}=\frac{L}{\ell}$, according to
Eq.~(\ref{difftun}). Our choice of units is motivated by the fact that
magnetic length $\ell$ and Coulomb energy $e^2/(\epsilon\ell)$ are
characteristic scales in the fractional--quantum--Hall regime. We
assumed $e^2 v_{\text{t}}/(\epsilon\hbar v_+ v_-)=0.1$ here but results
for different values can be obtained by linear scaling of
$V_{\text{eff}}$ or $L_{\text{eff}}$.
\label{condplot}}
\end{figure}
Solving Eq.~(\ref{currentequ}) in conjunction with Eq.~(\ref{fundament})
yields $I_{\text{J}}^{(4)}$ as a function of the externally applied
voltage $V=(\mu_1-\mu_3)/e$ and, hence, solves the transport problem
through line junctions having effective filling factor $\tilde\nu=1/4$.
While such a solution is accessible, even if only numerically, in the
general case, it is instructive to look at the special situation where
$\Delta=0$ and tunneling is weak. In this limit, we find for the
differential tunneling conductance the expression
\begin{equation}\label{difftun}
e \frac{d I_{\text{J}}^{(4)}}{d V}= \frac{e^2}{2\pi\hbar}\,\frac{1}{4}
\left\{1 - \cos\left(\frac{v_{\text{t}} L}{\hbar v_+ v_-}\, e V \right)
\right\}\quad ,
\end{equation}
which is valid as long as the expression in curly brackets is smaller
than unity. It exhibits an oscillatory dependence on both junction
length and applied voltage, which we illustrate in Fig.~\ref{condplot}
for a realistic set of parameters. The conductance oscillations as a
function of $V$ are a direct consequence of the velocity splitting of
the Majorana pseudofermions that are the quasiparticle excitations in
the line junction. This signature behavior for $\tilde\nu=1/4$ will
survive in the general case as long as the junction is narrow enough
such that $\Delta<v_{\text{t}}\ell^2|e V|/(\hbar v_{\text{b}}^2)$. Its
observation in real samples would provide strong evidence for the
reality of the Majorana quasiparticles in QH line junctions, showing yet
another example of how strongly--correlated condensed--matter systems
exhibit features of relativistic quantum--field theories typically
encountered in the realm of elementary--particle physics.

\section{Summary and conclusions}
\label{conclude}

In this article, we provide a detailed theory of transport through line
junctions formed between tunnel--coupled counter--propagating
fractional--quantum--Hall edge channels having different filling factors
of the Laughlin type. Strong correlations within the tunneling region
result in a nontrivial spectrum of elementary excitations when the
effective filling factor $\tilde\nu$ of the junction is equal to $1/2$
or $1/4$. An example for the former (latter) case is a line junction
between edge channels having filling factor 1 and $1/3$ ($1/5$).
Using bosonization and refermionization techniques, we map the original
line--junction problem to a model of tunneling between noninteracting
chiral spin--1/2 Dirac pseudofermions (for $\tilde\nu=1/2$) or a chiral
spinless p--wave superconductor  with Majorana--pseudofermion
quasiparticles (for $\tilde\nu=1/4$). In both situations, we find the
excitation spectrum and eigenstates of the line junction exactly.
Together with an exact general relation, given by Eq.~(\ref{fundament}),
between the externally applied chemical--potential difference $\mu_1-
\mu_3$ and intrinsic line--junction quantities, such as the total
current $I_{\text{J}}$, we can solve for the transport I--V
characteristics. The full solution for the case $\tilde\nu=1/2$ is
given by Eq.~(\ref{fullsol2}), together with Eq.~(\ref{transcoeff}). It
exhibits oscillations as a function of junction length that are similar
to those found in tunnel--coupled quantum wires~\cite{dircoupl1,
pank:jpc:00,uz:prb:01}. To calculate transport through junctions having 
$\tilde\nu=1/4$ in the most general case requires simultaneous numerical
solution of Eqs.~(\ref{currentequ}) and (\ref{fundament}) for specific
parameters realized in experiment. For the special situation of a narrow
junction, analytical results are available. We give the differential
tunneling conductance in Eq.~(\ref{difftun}), which shows oscillations
as a function of transport voltage. This behavior is the signature of a
nontrivial velocity splitting similar to spin--charge separation in
interacting quasi--1D systems. Its observation would provide evidence of
yet another mechanism for electron fractionalization in strongly
correlated electron systems. Experiments in samples of 90--degree
bent~\cite{mattcorner1,mattcorner2} or laterally
separated~\cite{kang:nat:99} 2D electron systems, suitably modified to
reach the fractional quantum Hall regime, could be used to test our
predictions.

\begin{acknowledgments}

Our work was supported in part by German Science Foundation (DFG) Grant
No.\ ZU~116/1 and a German--Israeli Project Cooperation (DIP) of the
German Ministry of Education and Research (BMBF). We enjoyed useful
discussions with N. Andrei, C.~de~C.~Chamon, P. Coleman, 
M.~Grayson, V. Oganesyan, F.~v.~Oppen, Y.~Oreg, S. L. Sondhi and
A.~Stern. E. S. wishes to thank the Department of Physics and Astronomy
at Rutgers University for the hospitality during part of the time this
work was carried out.

\end{acknowledgments}


\end{document}